\documentclass[%
reprint,
amsmath,amssymb,
aps,
pra,
superscriptaddress,
longbibliography,
]{revtex4-1}

\usepackage{graphicx}
\usepackage{dcolumn}
\usepackage{hyperref}

\usepackage{natbib}
\usepackage{color}
\usepackage{tabularx}
\usepackage{placeins}
\usepackage{enumitem}
\usepackage{braket}
\usepackage{upgreek}
\usepackage{acronym}
\usepackage[textsize=tiny, backgroundcolor=red!40, linecolor=red]{todonotes}

\begin{document}

\title{A silicon-integrated telecom photon-spin interface}

\author{L. Bergeron}
\affiliation{Department of Physics, Simon Fraser University, Burnaby, British Columbia, Canada V5A 1S6}
\author{C. Chartrand}
\affiliation{Department of Physics, Simon Fraser University, Burnaby, British Columbia, Canada V5A 1S6}
\author{A. T. K. Kurkjian}
\affiliation{Department of Physics, Simon Fraser University, Burnaby, British Columbia, Canada V5A 1S6}
\author{K. J. Morse}
\affiliation{Department of Physics, Simon Fraser University, Burnaby, British Columbia, Canada V5A 1S6}

\author{H. Riemann}
\affiliation{Leibniz-Institut f\"ur Kristallz\"uchtung, 12489 Berlin, Germany}
\author{N.~V.~Abrosimov}
\affiliation{Leibniz-Institut f\"ur Kristallz\"uchtung, 12489 Berlin, Germany}

\author{P. Becker}
\affiliation{Physikalisch-Technische Bundestanstalt Braunschweig, 38116 Braunschweig, Germany}

\author{H.-J. Pohl}
\affiliation{VITCON Projectconsult GmbH, 07743 Jena, Germany}

\author{M. L. W. Thewalt}
\affiliation{Department of Physics, Simon Fraser University, Burnaby, British Columbia, Canada V5A 1S6}
\author{S. Simmons}
\thanks{s.simmons@sfu.ca}
\affiliation{Department of Physics, Simon Fraser University, Burnaby, British Columbia, Canada V5A 1S6}

\date{\today}

\setlength{\skip\footins}{0.5cm}
\setitemize{topsep=0pt,parsep=0pt,partopsep=0pt}

\begin{abstract} Long-distance entanglement distribution is a vital capability for quantum technologies. An outstanding practical milestone towards this aim is the identification of a suitable matter-photon interface which possesses, simultaneously, long coherence lifetimes and efficient telecommunications-band optical access. In this work, alongside its sister publication \cite{SisterPRB}, we report upon the T center, a silicon defect with spin-selective optical transitions at 1326~nm in the telecommunications O-band. Here we show that the T center in $^{28}$Si offers electron and nuclear spin lifetimes beyond a millisecond and second respectively, as well as optical lifetimes of 0.94(1) $\upmu$s and a Debye-Waller factor of 0.23(1). This work represents a significant step towards coherent photonic interconnects between long-lived silicon spins, spin-entangled telecom single-photon emitters, and spin-dependent silicon-integrated photonic nonlinearities for future global quantum technologies.\end{abstract}

\maketitle

The global search for a high-performance quantum interface between telecom photons and long-lived matter qubits is ongoing \cite{Awschalom2018}. 
The predominant photon-spin candidates presently under study either do not operate at telecom wavelengths or are not hosted within silicon. 
Technologies such as frequency conversion \cite{Bock2018} or evanescent integration with silicon photonics \cite{Raha2020,Covey2019} are being developed to address the individual shortcomings of such interfaces.

Silicon is a convenient and attractive host for a photon-spin interface as it underpins both the most established integrated electronics and integrated photonics platforms. 
Silicon, and in particular isotopically purified $^{28}$Si, is a host to many atomically reproducible defects with exceptional spin and/or optical properties. 
Phosphorus donors possess ultra-long spin lifetimes of up to 3\,hours \cite{Saeedi2013} but do not interact strongly with light. 
Singly-ionized chalcogen donors, such as $^{77}$Se$^+$, offer both long-lived spins and relatively strong interactions with light \cite{Morse2017, DeAbreu2019}, however the wavelengths involved are in the technically challenging mid-infrared. 
Erbium defects in silicon \cite{Yin2013, Weiss2020} offer weak dipole-forbidden telecom optical transitions and potentially long-lived spins \cite{Hughes2019}, yet complex readily with other silicon defects into a wide selection of symmetry sites and complexes, only a small fraction of which are optically active \cite{Kenyon2005}. 
The family of silicon defects known as radiation damage centers, including the well-studied G, C, and W centers \cite{Chartrand2018, Beaufils2018, Redjem2020, Buckley2017, Tait2020}, emit photons in or near the telecommunications bands, but do not possess an unpaired electron spin in their optical ground states \cite{Chartrand2018}. 

A few less well-studied radiation damage centers, in particular the 
T \cite{Minaev1981, Irion1985, Safonov1996, Leary1998, Davies2006, Hayama2004, Henry1991,  Lightowlers1994hydrogen, Gower1997, Safonov1999, Lightowlers1994luminescence, Lightowlers1997, Safonov1994, Schmidt2000, Hayama2004}, I \cite{Henry1991, Lightowlers1994hydrogen, Gower1997, Safonov1999}, and M \cite{Safonov1999, Lightowlers1994luminescence, Lightowlers1997, Safonov1994, Schmidt2000} centers, were previously reported to have emission in the telecommunications bands which split under a magnetic field, and their ground states were believed to possibly have an unpaired electron spin. 
However, the most important photon-spin interface properties of these centers were unknown. For example, conflicting models of the T center, the most studied of the three, identified the ground state as either paramagnetic \cite{Safonov1996} or diamagnetic \cite{Irion1985}. 

\begin{figure}[]
\includegraphics[width=\linewidth]{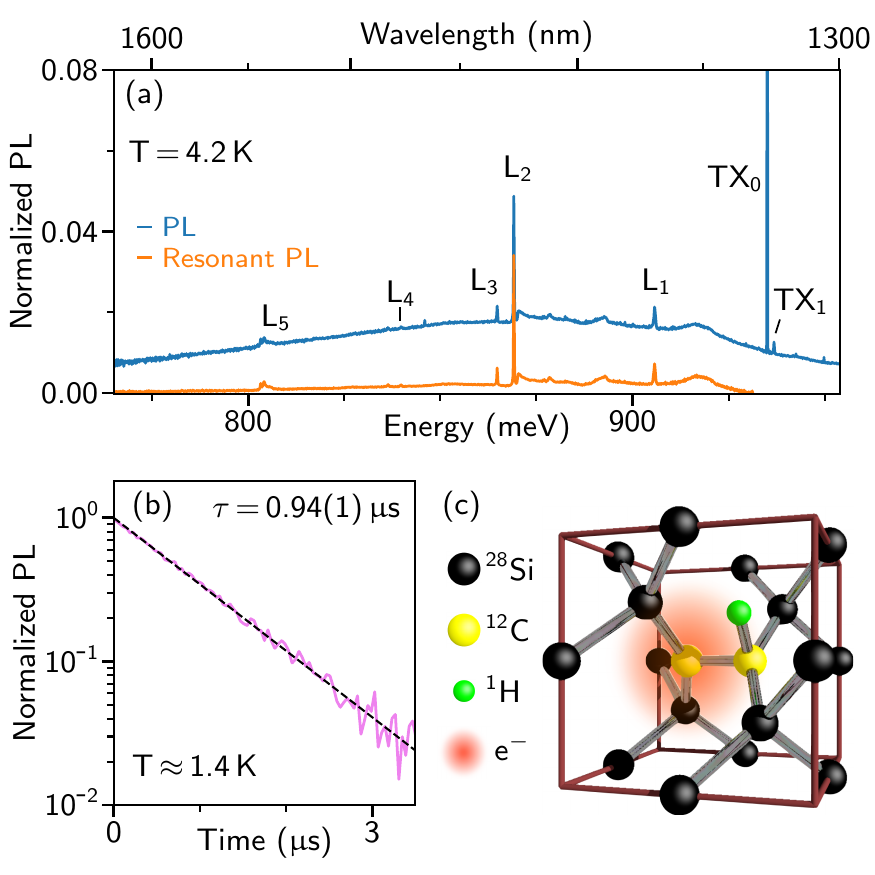}
\caption{Optical and luminescence properties of the T center photon-spin interface. (a) The photoluminescence (PL) spectrum with above-bandgap excitation (1047 nm, top trace) normalized to a T line (TX$_0$) intensity of unity, and a PL spectrum resonantly excited by pumping TX$_0$ (bottom trace), normalized to equal L$_2$ local vibrational mode intensity in both spectra. Together this data is used to extract a Debye-Waller factor of 0.23(1). (b) Lifetime measurements of the TX$_0$ BE. (c) Atomic structure of the T center as proposed by Ref. \cite{Safonov1996}.}
\label{fig:1}
\end{figure}

Some properties of the T center are available in the literature. 
Through isotope shift studies, T centers are known to be made of at least one hydrogen atom and two bonded carbon atoms \cite{Safonov1996, Leary1998}. 
The optically excited state, the ground state of a bound exciton (BE), is known to be a doublet split by about 1.8\,meV~\cite{Irion1985,Safonov1996}, and is known to thermally disassociate around 40\,K \cite{Irion1985} corresponding to a binding energy of around 30--35 meV \cite{Safonov1999}. 
Various components of the photoluminescence (PL) spectrum, including the two zero phonon line (ZPL) transitions and their local vibrational mode replicas, have been given labels as shown in Fig.~\ref{fig:1}(a). 
Detailed modelling of the atomic structure, the results of which are shown schematically in Fig.~\ref{fig:1}(c), and some speculation as to formation mechanisms of the T center were undertaken in Refs. \cite{Safonov1996,Leary1998,Davies2006}. 
A first look at ZPL shifts and temperature shifts of radiation damage centers in isotopically purified $^{30}$Si was undertaken in Ref.~\cite{Hayama2004}. 
The ZPL is known to split under the application of a magnetic field, revealing anisotropic g-values for the hole spin and isotropic g-values for the electron spin. 
Beyond these aforementioned results, relatively little was known about the T center prior to this work. 
In the model of Ref. \cite{Irion1985}, both the hole spin and uncoupled electron spin are in the BE state. In the model of Ref. \cite{Safonov1996} the uncoupled electron spin is a property of the ground state of the center, while the hole spin is a property of the BE state. 
Here, and in Ref.~\cite{SisterPRB}, we resolve this ambiguity in favor of Ref.~\cite{Safonov1996} and conclude that the ground state consists of an unpaired electron spin and the BE state consists of an electron singlet pair and an anisotropic hole spin. 

Isotopically purified silicon improves upon the properties of photon-spin interfaces in multiple ways. 
For one, the coherence lifetimes of the spins are increased by removing the $^{29}$Si nuclear spins as a source of dynamic magnetic noise \cite{Saeedi2013}. 
Secondly, the inhomogeneous line broadening of optical transitions is reduced by removing static local bandgap and binding energy variations which arise from the silicon isotope mass mixture inherent in natural silicon \cite{Karaiskaj2001}. 
Accordingly, to obtain the results in this work, the sample under investigation is a piece of isotopically purified $^{28}$Si crystal obtained from the Avogadro project \cite{Becker2010} which is enriched to 99.995\% $^{28}$Si, with $<10^{14}$\,O/cm$^3$ and $<5\times 10^{14}$\,C/cm$^3$. 
An ensemble of T centers was fabricated within this sample by 10\,MeV electron irradiation followed by an annealing recipe in stages up to 450\,$^\circ$C as described in Ref.~\cite{SisterPRB}.

This work demonstrates the T center's competitive optical properties: a Debye-Waller factor of 0.23(1) and a 0.94(1)~$\upmu$s emitter lifetime in the telecommunications O--band near 1326\,nm, very close to the zero dispersion wavelength of standard silica single-mode fibers. This work also demonstrates the T center's competitive spin properties, including Hahn-echo $T_2$ lifetimes beyond 1\,second. 

Detailed supporting results in Ref. \cite{SisterPRB} include: the isotope and temperature effects on ZPL linewidth and position; the BE's acceptor-like excited states; magneto-PL studies; continuous-wave optically detected magnetic resonance; as well as electron and nuclear spin initialization, readout, and Rabi oscillations. 

{\textit{Section A: Optical Properties.}---Photoluminescence (PL) from T centers can be observed at cryogenic temperatures using either nonresonant above-bandgap excitation, or resonant below-bandgap excitation. 
The application of above-bandgap light (in this instance 1047\,nm) generates free carriers and free excitons (electron-hole pairs) which localize onto defects, including T centers, forming BEs which can recombine radiatively. 
In Fig.~\ref{fig:1}(a) the resulting emitted light is characterized using a Bruker IFS 125 HR spectrometer, a CaF$_2$ beamsplitter, and a germanium diode detector (see \cite{SisterPRB}). 
This spectrum agrees with previous T center PL spectra in the literature \cite{Irion1985}. 
Here we label a few features of interest including the ZPL doublet (The BE `TX' level is split into TX$_0$ and TX$_1$) and local vibrational mode (LVM) replicas (L$_i$, after Ref.~\cite{Safonov1996}). 
The substantially narrower zero phonon optical lines available using a $^{28}$Si sample yield an estimate of the TX level splitting of 1.76(1) meV at 4.2~K.

A proportion of the light emitted by T centers is emitted in the ZPL, and the rest is produced in the phonon sideband: the ZPL proportion is given by the Debye-Waller factor \cite{Aharonovich2011}. 
Other luminescent defects in the sample may also contribute to the observed PL signal, as can clearly be seen from the broad PL at energies both higher and lower than the TX$_0$ ZPL. 
In order to measure the Debye-Waller factor accurately, we suppress this non T-related PL by employing resonant PL spectroscopy. 
A Toptica DL100 laser drives the TX$_0$ ZPL resonantly. The resulting emitted light passes through appropriate long-pass filters (1330\,nm) to block the laser wavelength and is then routed to the Fourier transform infrared (FTIR) spectrometer for characterization. 
The resulting resonant PL phonon sideband spectra, taken with the same resolution as the matching above-bandgap PL data, and with the silicon laser Raman line subtracted, is shown in Fig.~\ref{fig:1}(a). 
The sharp LVM feature L$_2$ was used to normalize the two PL traces so that an accurate Debye-Waller factor of 0.23(1) could be determined.

The lifetime of the TX level can be measured by applying pulsed above-bandgap excitation and measuring the luminescence decay of light emitted into the ZPL line. 
In this instance, 965~nm light pulses, at a repetition rate of 125 kHz and with sub-ns duration, were generated by a Picoquant PLD 800-D and directed to the sample held at a temperature of either 1.4\,K or 4.2\,K. 
The T ZPL emission was filtered through a double spectrometer and routed into an ID230 single-photon detector whose output was directed to a multi-channel scaler with 40\,ns time resolution. 
This data is shown in Fig.~\ref{fig:1}(b) and reveals a TX$_0$ BE lifetime of 0.94(1)\,$\upmu$s. 
Repeating these lifetime measurements using a range of silicon samples with different defect concentrations at  either 1.4\,K or 4.2\,K reveals the same lifetime values within the margin of error. 
This increases our confidence that this reported lifetime is inherent to the T center and is not due to, for example, the free-exciton decay time. 
These excited state lifetimes are sufficiently long to consider the excitonic degree of freedom as an additional local quantum resource. 

The radiative quantum efficiency of this optical transition is presently unknown. 
Silicon's maximum phonon energy of $\sim\!65$~meV implies that pure non-radiative decay would require the simultaneous emission of at least 15 phonons. 
We were unable to observe any sample conductivity changes due to non-radiative Auger recombination processes, which is a highly successful method for measuring the nonradiative BE decay of low-concentration phosphorus donors in silicon \cite{Saeedi2013}. 
Working under the assumption that the recombination lifetime is entirely from radiative processes, this BE lifetime corresponds to a total transition dipole moment of 1.52(1) Debye and a ZPL transition dipole moment of 0.73(2) Debye. 
From single-center linewidth upper bounds of 33(2)~MHz as presented in \cite{SisterPRB}, the above ZPL transition dipole moment would give a single-defect cooperativity of 1 in a photonic cavity with a Q-factor of $1\times10^4$ with a realistic mode volume of $(\lambda/n)^3$.

\begin{figure}[h!]
\includegraphics[width=\linewidth]{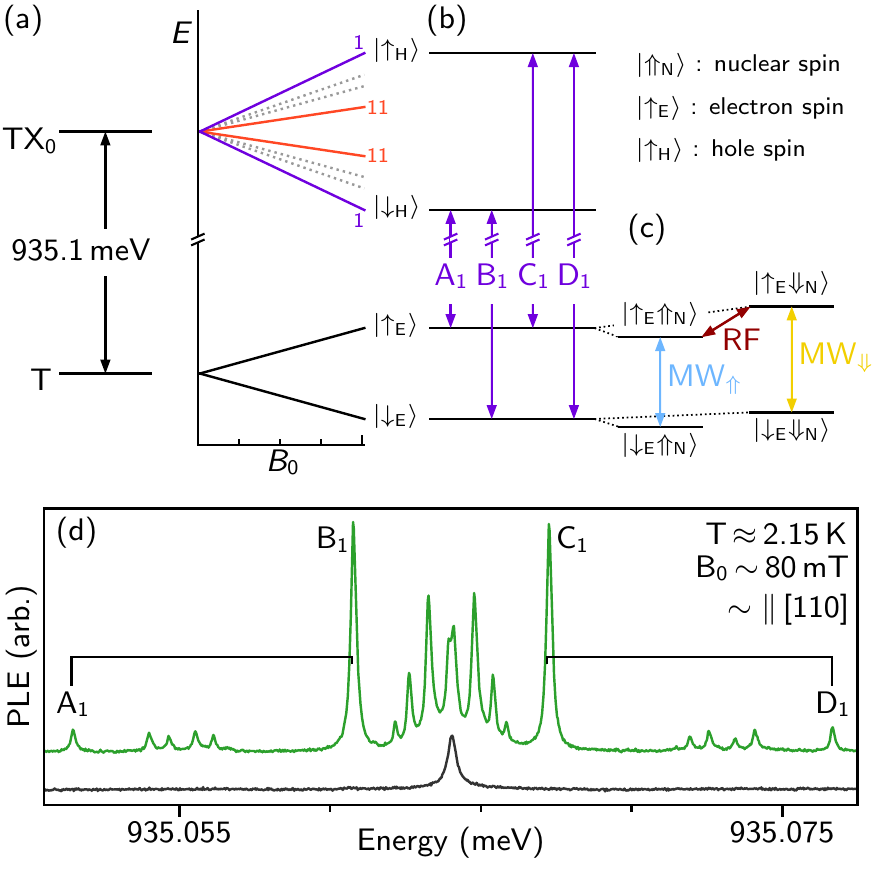}
\caption{ (a) Level structure and magnetic field dependence of the ground and lowest energy BE state of the T center. The anisotropic unpaired hole spin in TX$_0$ reveals 11 distinct orientational subsets. 
(b) The optical transitions A$_1$, B$_1$, C$_1$, D$_1$ correspond to those available to a single orientational subset, here subset 1. 
(c) Schematic of the spin levels and transitions of subset 1 under our experimental conditions. The nuclear spin states are not optically resolvable. 
(d) PLE of the observed TX$_0$ ZPL transitions in the presence of a static magnetic field. Top (green): PLE spectrum using the depolarizing scheme for subset 1 as discussed in the text. Bottom (black): PLE spectrum without the application of any depolarizing magnetic resonance.  The black brackets joining A$_1$ with B$_1$, and C$_1$ with D$_1$, show the $g_\textrm{E}=2.005$ isotropic electron Zeeman splitting.}
\label{fig:2}
\end{figure}

{\textit{Section B: Spin Properties.}---Next we turn to the T center spin degrees of freedom. 
We first report methods for optically addressing an orientation-specific subset of T centers, then follow with the process for spin initialization and measurement of that subset, and conclude with the measurement of spin $T_1$ lifetimes and Hahn-echo $T_2$ times for both the hydrogen nuclear spin and unpaired electron spin.

The energy level diagram of the T center photon-spin interface is shown in Fig.~\ref{fig:2}(a--c). 
The ground T level consists of a single unpaired electron, as well as up to three hyperfine-coupled nuclear spins.
The TX levels consist of an additional exciton: the two electrons form a singlet and the remaining unpaired spin-3/2 hole determines the magnetic properties of the TX states, which are split into TX$_0$ and TX$_1$ due to the symmetry-lowering strain field of the defect. 
For a given T center in a magnetic field this produces a quartet of spin-selective optical transitions between the T state's isotropic electron spin levels and the TX$_0$ state's anisotropic hole spin levels (Fig.~\ref{fig:2}b).
In this work, the hyperfine substructure is not optically resolvable.
An applied magnetic field was known to split the ensemble TX$_0$ ZPL transition into at least six distinct quartets of optical transitions corresponding to at least six orientations of the T center relative to the direction of the applied magnetic field \cite{Irion1985}.
The reduction of inhomogeneous broadening in the $^{28}$Si samples studied here allows for the observation of even more optical structure: 11 orientational subsets are optically resolved in \cite{SisterPRB}, as indicated in Fig \ref{fig:2}(a). 
From the proposed atomic structure of the center shown in Fig.~\ref{fig:1}(c) we would expect to see up to 12 distinct orientational subsets under an applied magnetic field \cite{Kaplyanskii1967}.

The ground state T level spin Hamiltonian $\mathcal{H_T}$ with one $^{1}$H and two $^{12}$C constituents is given by 
\begin{equation}
\mathcal{H_T} = \mu_\textrm{B} {\bf B_0 g_E  S} + \mu_\textrm{N} g_\textrm{N} {\bf B_0 I} + h {\bf S  A I} \label{eq:ham}
\end{equation}
where $\mu_\textrm{B}$ is the Bohr magneton, ${\bf B_0}$ is the magnetic field vector, ${\bf g_E}$ is the electron spin g factor tensor which is approximately isotropic with ${g_\textrm{E} = 2.005(8)}$, $\bf{S}$ is the electron spin vector, $\mu_\textrm{N}$ is the nuclear spin magneton, ${g_\textrm{N}}$ is the hydrogen nuclear spin g factor, $\bf{I}$ is the hydrogen nuclear spin vector, $h$ is the Planck constant, and $\bf{A}$ is the hyperfine tensor. 
For the purposes of this work we only consider the Zeeman term $\mu_\textrm{B} {\bf B_0 g_H  H}$ of the hole spin Hamiltonian for the TX$_0$ level, where ${\bf g_H}$ is the hole spin g factor tensor and $\bf{H}$ is the TX$_0$ hole spin vector.

In agreement with Ref.~\cite{Safonov1999} we find ${\bf g_E}$ to be almost entirely isotropic. 
The highly anisotropic g factor of the TX$_0$ BE hole spin, ${\bf g_H}$, thus allows for the optical selection of an individual orientational subset of T centers in a relatively small magnetic field. 
The quartet of transitions labelled A$_1$, B$_1$, C$_1$, D$_1$  in Fig.~\ref{fig:2}(b) corresponds to a particular orientational subset that is spectrally distinct from the others, which we will hereafter refer to as subset 1. 
These transitions can be driven resonantly by sweeping the frequency of a tunable single-frequency laser, and the resulting emitted light can be filtered and detected in a photoluminescence excitation (PLE) measurement scheme as described in \cite{SisterPRB}. 
The ground state spin level diagram for orientational subset 1 in a magnetic field of \mbox{80~mT} applied parallel to [110] is shown in Fig.~\ref{fig:2}(c). 
The electron and nuclear spin transitions can also be driven resonantly using magnetic resonance as described in \cite{SisterPRB}. 
The microwave (MW) and radio-frequency (RF) transitions we used in this work are shown in Fig.~\ref{fig:2}(c).

In the absence of any applied magnetic resonance signals, in a magnetic field strength $| {\bf B_0} | \sim 80$~mT applied roughly parallel to [110], the PLE spectrum generated by sweeping the laser energy near the ZPL energy consists of a single peak as shown in the bottom (black) trace of Fig.~\ref{fig:2}(d). 
This central peak is substantially weaker than the same experiment performed with no applied magnetic field. 
It is the signal generated by the orientational subset(s) where the B$_i$ and C$_i$ optical transitions remain nearly degenerate because the effective hole g factor is close to that of the ground state electron for the chosen direction of ${\bf B_0}$. 
All other orientational subsets have their electron spin efficiently hyperpolarized \cite{Steger2011} over the course of the laser energy sweep and do not generate enough photons during this process to be detected in PLE.

In PL up to 11 orientational subsets of T centers are optically resolvable~\cite{SisterPRB}.
To directly observe more of the orientational subsets using PLE in an applied magnetic field some degree of electron spin depolarization is required. 
When two MW frequencies centered about the isotropic $g=2.005$ electron spin resonance frequency are applied, here differing in value by $-$2.9\,MHz matching the effective hydrogen hyperfine value of orientational subset 1 in the chosen magnetic field, the electron spin is optimally depolarized. 
This results in a recovered PLE signal from subset 1, as well as some PLE signal from other subsets, as seen in the top (green) trace of Fig.~\ref{fig:2}(d). 
In Ref.~\cite{SisterPRB} we explain how this negative hyperfine value was obtained. 
For the remainder of the work described in this text the laser energy is set to the optical transition B$_1$ as shown in Fig.~\ref{fig:2}(d), for the purposes of initializing and measuring the spins associated with orientational subset 1. 

As depicted in Fig.~\ref{fig:2}(b, c), the chosen laser energy excites both nuclear spin configurations associated with the electron $\ket{\downarrow_\textrm{E}}$ state. 
Because of the long spin relaxation times in the ground T level, in the absence of any externally-driven spin mixing or depolarizing processes the spins efficiently hyperpolarize into the electron spin $\ket{\uparrow_\textrm{E}}$ state. 
To initialize a specific electron-nuclear spin eigenstate, we apply a laser resonant with the B$_1$ transition as well as an electron spin mixing scheme with a single MW frequency conditional upon the nuclear spin $\ket{\Downarrow_\textrm{N}}$ state, as shown by the MW$_\Downarrow$ transition in Fig.~\ref{fig:2}(c). 
After a time the spins hyperpolarize into the state $\ket{\uparrow_\textrm{E}\Uparrow_\textrm{N}}$. 
This is the procedure referred to as `POL' in Fig.~\ref{fig:3}. 

\begin{figure}[]
    \includegraphics[width=\linewidth]{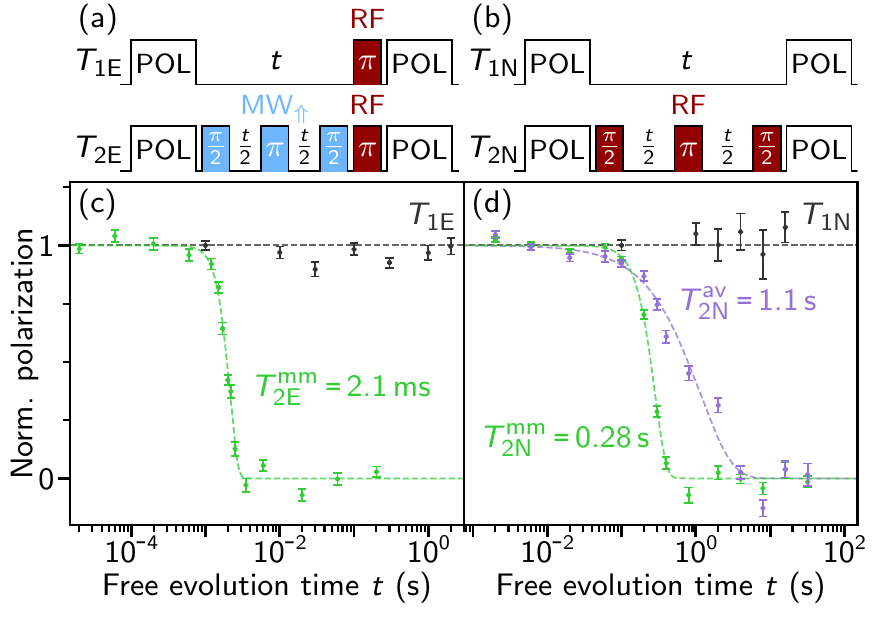}
    \caption{
        (a) and (b) Pulse sequences used to measure the electron-spin (a) and nuclear-spin (b) $T_1$ and $T_2$ times. The pulse colors match the spin transition frequencies in Fig.~\ref{fig:2}(c). The initialization and measurement POL protocols are common to all four sequences. 
       (c) and (d) Normalized $T_1$ and Hahn-echo $T_2$ decay curves for the electron (c) and hydrogen nuclear spin (d) for one orientational subset of T centers.  The electron spin Hahn-echo $T^{\textrm{av}}_{2\textrm{E}}$ time is limited by phase noise beyond 2\,ms. The hydrogen nuclear spin Hahn-echo $T^{\textrm{av}}_{2\textrm{N}}$  time is limited by phase noise beyond  280\,ms, which can be partially circumvented by using maximum magnitude techniques (see text), revealing $T^{\textrm{mm}}_{2\textrm{N}}>  1$\, second. }
    \label{fig:3} 
\end{figure}

The spin measurement procedure uses a similar technique. 
The electron spin optical hyperpolarization process in the absence of spin mixing is too efficient to generate a large number of photons for detection, however the nuclear spin hyperpolarization process using conditional MW mixing is sufficiently slow to detect a luminescence transient, as shown in Ref.~\cite{SisterPRB}. 
By integrating the luminescence transient while applying the POL procedure we can measure the nuclear spin $I_Z$ observable. 
In between initialization/measurement POL sequences the laser was blocked with a mechanical shutter. 

Resonant MW and RF pulses are used to extract the $T_1$ and Hahn-echo $T_2$ times of both the electron spin and hydrogen nuclear spin, as shown in Fig.~\ref{fig:3}(a) and \ref{fig:3}(b). 
In the case of the electron spin $T_1$ and $T_2$ data, the electron spin $S_Z$ observable is mapped to the nuclear spin $I_Z$ observable for readout using a leading conditional $\pi$ pulse (a CNOT gate). 
The signal to noise ratio on a single shot was just above 1 in the case of the nuclear spin observable, and below 1 for the electron spin, due to imperfect pulses generated by our home-built magnetic resonance assembly. 
As such, averaging was required to extract accurate lifetimes and coherence times. 

The resulting normalized $T_1$ lifetimes and $T_2$ Hahn-echo coherence times of the electron spin and nuclear spin are shown in Fig.~\ref{fig:3}(c) and \ref{fig:3}(d), respectively. 
The $T_1$ times for both the electron spin ($T_{1\textrm{E}}$) and nuclear spin ($T_{1\textrm{N}}$) are far beyond their measured Hahn-echo times; we observed no signal decay out to 16 seconds for both the electron spin and nuclear spin. 
Averaged Hahn-echo times, extracted by fitting the averaged data to stretched exponentials, are 2.1(1)\,ms (stretch factor 4.1(7)) and 0.28(1)\,s (stretch factor 2.9(4)) for the electron spin ($T^{\textrm{av}}_{2\textrm{E}}$) and nuclear spin ($T^{\textrm{av}}_{2\textrm{N}}$) respectively. 
These high stretch factors are consistent with instrumental phase noise as has been observed in similar spin measurements \cite{Saeedi2013}. 
In the case of the nuclear spin there was a sufficient signal to noise ratio to use maximum-magnitude techniques \cite{Saeedi2013} to partially mitigate this instrumental limitation. 
The top 10\% highest measurement values per time point were averaged to establish a tighter lower bound on the true Hahn-echo $T_2$, and this resulted in an exponential decay of $T^{\textrm{mm}}_{2\textrm{N}} = 1.1(2)$\,s (stretch factor 1). 
Moving forward, dynamical decoupling \cite{Saeedi2013} or clock transition \cite{Wolfowicz2012} techniques could be employed to extract even longer coherence times.

As described above, this study demonstrated the suitability of the T center as an integrated silicon telecom photon-spin interface. 
We reported sub-$\upmu$s optical lifetimes at telecom wavelengths, with no observable Auger nonradiative recombination, and a Debye-Waller factor of 0.23(1).
Furthermore, we performed optically detected magnetic resonance upon the electron and hyperfine-coupled hydrogen nuclear spin, and extracted competitive Hahn-echo $T_2$ spin lifetimes for each. 
Future studies incorporating $^{13}$C nuclei will allow for the investigation of T centers with four spins, which may prove useful for modular quantum information processing, error detection, and quantum state purification objectives. 
These results pave the way for the construction of new hybrid photon-spin all-silicon quantum integrated devices suitable for distributed quantum computing and global quantum networking.

This work was supported by the Natural Sciences and Engineering Research Council of Canada (NSERC), the Canada Research Chairs program (CRC), the Canada Foundation for Innovation (CFI), the B.C. Knowledge Development Fund (BCKDF), and the Canadian Institute for Advanced Research (CIFAR) Quantum Information Science program. 
The $^{28}$Si samples used in this study were prepared from the Avo28 crystal produced by the International Avogadro Coordination (IAC) Project (2004-2011) in cooperation among the BIPM, the INRIM (Italy), the IRMM (EU), the NMIA (Australia), the NMIJ (Japan), the NPL (UK), and the PTB (Germany). 
We thank Alex English of Iotron Industries for assistance with the electron irradiations.

\FloatBarrier
\bibliographystyle{apsrev4-1}

%

\end{document}